\documentclass[preprint,times]{elsarticle}
\usepackage{amsmath}
\usepackage{graphicx}
\usepackage{color}
\usepackage{CJK}
\usepackage{hyperref}
\usepackage{soul}
\usepackage{geometry}

\usepackage{setspace}

\journal{***}

\begin{document}
	\begin{frontmatter}
		
		\title{Electrostatic self$-$assembly of neutral particles on a dielectric substrate: A theoretical study via a multiple$-$image method and an effective$-$dipole approach}

		\author{Xin Li} \author{Changhao Li} \author{Xiangui Chen} \author{Zaixin Wang} \author{Sun Min}
		
		\author{Decai Huang \corref{Huang}}
		\cortext[Huang]{hdc@njust.edu.cn, 086$-$15951864599}
		\address{Department of Applied Physics, Nanjing University of Science and Technology, Nanjing 210094, China}
		\address{Engineering Research Center of Semiconductor Device Optoelectronic Hybrid Integration in Jiangsu Province} 
		
		\date{\today}
		
		\begin{abstract} %\('\)
			
			A multiple$-$image method is developed to accurately calculate the electrostatic interaction between neutral dielectric particles and a uniformly charged dielectric substrate. The difference in dielectric constants between the particle and the solvent medium leads to a reversal of polarization direction of the particle. The variance in dielectric constants between the solvent medium and the substrate causes a transition from attractive to repulsive forces between the particle and the substrate. A nonuniform electrostatic field is generated by the polarized charges on the substrate due to mutual induction. These characteristics of electrostatic manipulation determine whether particles are adsorbed onto the substrate or pushed away from it. The self$-$assembled particles tend to aggregate in a stable hexagonal structure on the substrate. These findings provide new insights into self-assembly processes involving neutral particles on a dielectric substrate.
			
		\end{abstract}
		
		
	\end{frontmatter}
	
	%\section*{Introduction}
	%\label{Introduction}
	
	Self-assembly of particles is a fundamental process in colloidal solution, wherein individual components autonomously arrange themselves into ordered structures and patterns. These self$-$assemblies have exhibited novel physical and chemical properties that differ from their bulk material counterparts, including unique characteristics in electronics, optical absorption, catalysis, mechanical rheology, and drug delivery\cite{Huang2020,Levay2018,Liang2014}. Substantial efforts have been dedicated to developing methods for self$-$assembled particles, encompassing metals, polymers, semiconductors, and neutral dielectric particles\cite{Bishop2009,Tang2016,Walker2011,Panat2022,Tabassian2016}. 
	The assembly strategies for particles are determined by their intrinsic properties and the environment, i.e., magnetism\cite{Rossi2021,Mann2019,Spatafora$-$Salazar2021,Ruiz$-$Lopez2017,Domingos2020}, electrostatics\cite{Yakovlev2022,Sherman2018,VanBlaaderen2013,Almudallal2011}, and pH value of solvent\cite{Cho2009} and external magnetic and electric fields, which influence the types of interparticle interactions. The primary objective of these methods is to adjust the attractive and repulsive interactions between the particles, thereby modulating both their strength and range.
	
	In a colloidal solution, suspended particles can aggregate together through various interparticle forces, such as hydrogen bonding and hydrophobic interactions, as well as externally driven forces such as light and magnetic fields. Electrostatic manipulation of self$-$assembly holds significant importance due to its controllability\cite{Bishop2009,Colla2018}.
	When particles carry net charges, such as ion particles and charge-decorated particles, an external electric field will propel them in a consistent direction\cite{Yoon2017}.
	However, when the particles are neutral, electrostatic manipulation of self$-$assembly becomes subtle, depending on the dielectric constants of the particle $\varepsilon_{p}$ and the solvent medium $\varepsilon_{m}$. 
	%{\color{red}\st{Under the condition of $\varepsilon_{p}>\varepsilon_{m}$, positive polarization occurs, aligning the polarization moments with the external electric field. Conversely, when $\varepsilon_{p}<\varepsilon_{m}$, the polarization moment reverses, leading to negative polarization.}} 
	Under the condition $\varepsilon_{p}>\varepsilon_{m}$, the polarization direction of particles aligns with the direction of the external electric field. Conversely, when $\varepsilon_{p}<\varepsilon_{m}$, the direction of the polarization direction reverses.
	Regardless of the type of polarization, polarized particles can gather together in a chain$-$like structure\cite{Cetin2011,Helal2016,Fertig2021}. 
	In cases where the external electrostatic field is non-uniform, such as around a non$-$planar electrode, spatial variations in the field gradient lead to different polarized charge distributions on each side of the particle. 
	The external nonuniform electric field creates a differential electrostatic force on both sides, resulting in dielectrophoresis, causing aggregation and separation of diverse particles on the substrate. 
	To control the packing behavior and functionality of assembled particles on the substrate, different substrates, such as metal and dielectric electrode with varying dielectric constants, are employed to manipulate the electrostatic field distribution around the substrate\cite{Mohammad2019,Huo2019,Woehl2014,Suehiro2019,Verveniotis2011}. Therefore, studying the influence of the substrate\('\)s properties on the electrostatic interaction between the particles and thus the bottom$-$up self-assembly process is a crucial research issue\cite{Lindgren2018a,Liu2002,Lee2011}.
	
	To quantitatively determine the electrostatic interaction between particles, experimental studies have directly investigated a bi$-$spherical system comprising two identical conducting or dielectric particles\cite{Cho2021,Rupp2018,Mittal2008,Chiu2014}. Wang et al. discovered that the interaction force $(F)$ be increases with the strength of external electric field $(E_0)$ and follows a quadratic relation of $F\propto{E_0^2}$\cite{Gao2015,Liu2015,GaoX2015}. Similar findings have been reported under DC and low$-$frequency AC electric fields for two metal$-$oxide spheres $\rm{SrTiO_{3}}$. Various theoretical methods have also been developed to analyze the electrostatic interaction between particles, such as finite element analysis (FEA) and multipole moment expansion {\it et al.}\cite{Bichoutskaia2010,Khachatourian2014,Lindgren2016,Lindgren2018b,Hassan2022,Meyer2015,Tao1995,Davis1992}. These foundational studies indicate that the polarization effect significantly influences the electrostatic interaction between neutral particles. 
	In our previous works, a novel approach known as the multiple$-$image method (MIM) was introduced, and the theoretical results using this method achieved good agreement with the experimental findings\cite{Gao2012,Li2022,Li2024}. In this study, the MIM is extended to systems involving a substrate. By incorporating MIM, the mutual dipolar inductions to include a substrate are taken into account, enabling the modulation of attractive and repulsive electrostatic interactions between the particles by adjusting the ratio of their dielectric constants.
	
	This paper is organized as follows. 
	%In Sec.\ref{Theory}, 
	Detailed introduction to the theoretical methods, including EDA and MIM, are presented for a system consisting of a neutral spherical particle and a uniformly charged substrate.
	%In Sec.\ref{Results}, 
	Two special cases, involving the attractive and repulsive electrostatic interactions between the particle and the substrate, are initially explored using both EDA and MIM. The analysis includes a study of the corresponding electric potential. A comprehensive tabulation summarizing the attractive and repulsive electrostatic interactions between the particles and the substrate is provided for various dielectric constants of particles, solvent medium, and substrates. 
	Additionally, the spatial distribution of interaction energy for a system with two particles is illustrated. The electrostatic interactions for multiple particles around the substrate are calculated, and conclusions regarding the stable packing structure are drawn.
	%In Section \ref{Conclusions}, 
	The findings of the study are summarized.
	
	\section*{THEORETICAL ARGUMENTS}
	\label{Theory}
	
	A planar dielectric substrate, assumed to be infinitely extended and uniformly charged, is always to serve as a source for a uniform electric field throughout the space. According to Gauss\('\)s law, the magnitude of this electric field is determined by:
	
	\begin{equation}
		{\mathop{E}\limits^{\rightharpoonup}}_0 = \frac{\sigma}{2\varepsilon_{m}\varepsilon_0},
		\label{eq:E1}
	\end{equation}
	where 
	\(\sigma\) represents the surface charge density on the substrate,
	\(\varepsilon_m\) and \(\varepsilon_0\) are the relative permittivity of the medium and the permittivity of vacuum, respectively.
	
	A spherical dielectric particle with radius $R$ will become polarized in a uniform electric field ${\mathop{E}\limits^{\rightharpoonup}}_0$. According to the Clausius$-$Mossotti (CM) relation, it is widely accepted that the induced charges on the particle can be effectively approximated by an equivalent electric dipole moment,
	\begin{equation}
		\mathop{p}\limits^{\rightharpoonup}=\frac{\varepsilon_{ p}-\varepsilon_{ m}}
		{\varepsilon_{ p}+2\varepsilon_{ m}}
		4\pi\varepsilon_{m}\varepsilon_{0} R^{3}{\mathop{E}\limits^{\rightharpoonup}}_0,
		\label{eq:p1}
	\end{equation}
	\noindent where $\varepsilon_{ p}$ is the relative permittivity of the particle\cite{Guo2008}.
	
\begin{figure*}[htbp]
		\centering
		\includegraphics[width=0.65\textwidth,trim=145 250 80 210,clip]{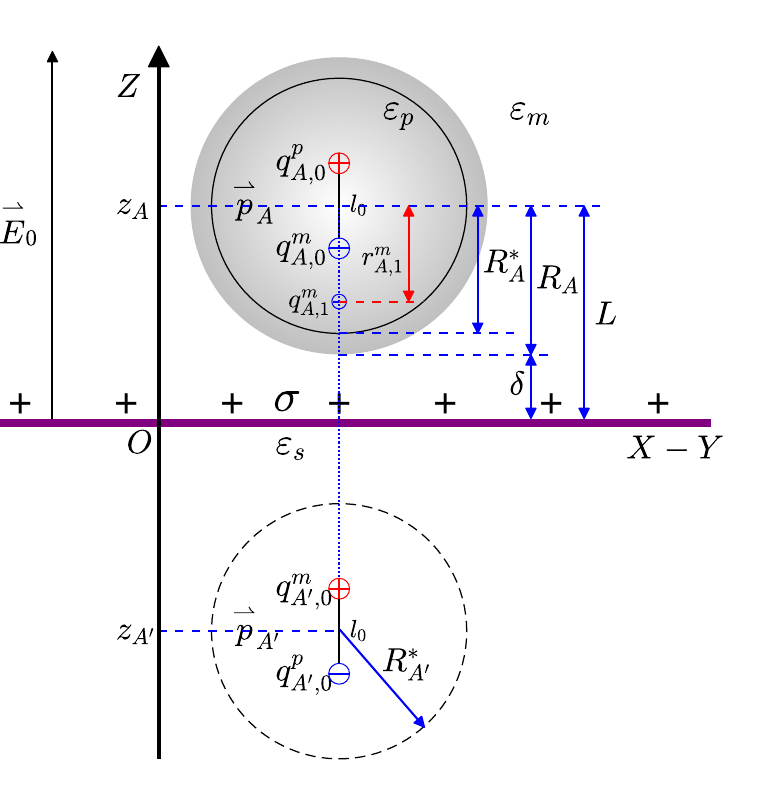}
		\caption{(color online) Schematic of electrostatic interaction between a dielectric particle with radius $R_A$ and an infinite planar substrate with uniform initial charge density $\sigma$.
		The relative permittivities of the particle, the planar substrate, and the medium are denoted as $\varepsilon_p$, $\varepsilon_s$, and $\varepsilon_m$ respectively. $L$ is the distance from the center of the dielectric particle to the planar substrate. 
		The hypothetical conducting particle and the imageed one are denoted in solid and dashed circular lines, respectively. Their radii are $R_{A}^*$ and $R_{A'}^*$, respectively. 	
		$q_{A,0}^{p}$ and $q_{A,0}^{m}$ are the initial polarized charges induced by external uniform electric field  ${\mathop{E}\limits^{\rightharpoonup}}_0$, separated by a distance $l_0$. $q_{A',0}^{m}$ and $q_{A',0}^{p}$ are the image charges of $q_{A,0}^{m}$ and $q_{A,0}^{p}$, respectively, with respect to the planar substrate. \(q_{A,1}^{m}\) is the image charge of \(q_{A',0}^{m}\) in particle $A$. $r_{A,1}^{m}$ represents the distance between $q_{A,1}^{m}$ and the center of particle $A$.}
		\label{fig:Fig1Model}
	\end{figure*}

	An infinitely extended planar substrate, characterized by a uniform surface charge density of \(\sigma\) and a dielectric particle with radius \(R_A\) are arranged as shown in Fig.\ref{fig:Fig1Model}. The center of the particle and the planar substrate are located at \(z_{A}\) and \(0\) on the z$-$axis, resulting in a separation distance of \(L\) between them. The separation gap between the lower surface of particle $A$ and the substrate surface is $\delta$. The relative permittivities of the particle, the planar substrate, and the medium are denoted as \(\varepsilon_p\), \(\varepsilon_s\), and \(\varepsilon_m\) respectively. In Fig.\ref{fig:Fig1Model}, we illustrate the case under the condition of $\varepsilon_p > \varepsilon_m$ and $\varepsilon_s > \varepsilon_m$. The mutual inductance between the particle and the planar substrate can be calculated through various analytical methods.
	
	\subsection*{EFFECTIVE$-$DIPOLE APPROACH} %Effective$-$Dipole Approach}
	\label{EDA}
	
	In our previous work\cite{Li2022,Li2024}, the effective$-$dipole approach (EDA), originally introduced by Chan et al.\cite{Chan2020,Guo2023,Zhang2023}, is extended to macroscopic dielectric particles. With this methodology, a dielectric particle in an external uniform electric field  ${\mathop{E}\limits^{\rightharpoonup}}_0$ can be treated as a hypothetical conducting particle with a reduced radius  ${R_{A}^{*}}^{3}=\rvert\frac{\varepsilon_{ p}-\varepsilon_{ m}}{\varepsilon_{ p}+2\varepsilon_{ m}}\rvert R_A^{3}$, enabling us to express Eq.(\ref{eq:p1}) as:
	
	\begin{equation}
		{\mathop{p}\limits^{\rightharpoonup}}_A=\frac{\varepsilon_{ p}-\varepsilon_{ m}}{\rvert\varepsilon_{ p}-\varepsilon_{ m}\rvert} 4\pi\varepsilon_{m}\varepsilon_{0}{R_A^{*}}^{3}{\mathop{E}\limits^{\rightharpoonup}}_0,
		\label{eq:p2}
	\end{equation}
	\noindent which indicates two reversed polarization conditions in a external uniform electric field. When $\varepsilon_{p} > \varepsilon_{m}$, dipole moment of ${\mathop{p}\limits^{\rightharpoonup}}_A$ aligns with the direction of ${\mathop{E}\limits^{\rightharpoonup}}_0$. Conversely, when $\varepsilon_{p} < \varepsilon_{m}$,  reversed polarization happens, in which the polarization of ${\mathop{p}\limits^{\rightharpoonup}}_A$ is opposite to that of ${\mathop{E}\limits^{\rightharpoonup}}_0$.
	
	To calculate the total polarization of particle $A$, ${\mathop{p}\limits^{\rightharpoonup}}_A$, two components need to be considered: the external electric field ${\mathop{E}\limits^{\rightharpoonup}}_0$ and the induced electric field generated by the dipole moment ${\mathop{p}\limits^{\rightharpoonup}}_{A'}$, which acts as the image dipole of ${\mathop{p}\limits^{\rightharpoonup}}_A$ on the planar substrate. For the latter, it is typically assumed to be uniform within particle $A$, with its magnitude equal to that at the center of the particle.
	
	\begin{subequations}
		\begin{align}
			&{\mathop{p}\limits^{\rightharpoonup}}_A=\frac{\varepsilon_{ p}-\varepsilon_{ m}}{\rvert\varepsilon_{ p}-\varepsilon_{ m}\rvert}
			4\pi\varepsilon_{m}\varepsilon_{0}{R_A^{*}}^{3}{\mathop{E}\limits^{\rightharpoonup}}_A,\label{eq:p4a}\\
			&{\mathop{E}\limits^{\rightharpoonup}}_A={\mathop{E}\limits^{\rightharpoonup}}_0+\frac{2{\mathop{p}\limits^{\rightharpoonup}}_{A'}}{4\pi\varepsilon_{m}\varepsilon_{0}(2L)^{3}},\label{eq:E4b}\\
			&{\mathop{p}\limits^{\rightharpoonup}}_{A'}=-\frac{\varepsilon _m-\varepsilon _s}{\varepsilon _m+\varepsilon _s} {\mathop{p}\limits^{\rightharpoonup}}_A,\label{eq:p4c}
		\end{align}
	\end{subequations}
	where $\frac{2{\mathop{p}\limits^{\rightharpoonup}}_{A'}}{4\pi\varepsilon_{m}\varepsilon_{0}(2L)^{3}}$ is the electric field of the image dipole ${\mathop{p}\limits^{\rightharpoonup}}_{A'}$ at the center of particle $A$. A detailed derivation of Eq.(\ref{eq:p4c}) can be found in \ref{Appendix A}. In Eq.(\ref{eq:p4c}), two interesting observations are made. Firstly, the magnitude of ${\mathop{p}\limits^{\rightharpoonup}}_{A'}$ is smaller than that of ${\mathop{p}\limits^{\rightharpoonup}}_A$. Secondly, the direction of ${\mathop{p}\limits^{\rightharpoonup}}_{A'}$ is related not only to the direction of ${\mathop{p}\limits^{\rightharpoonup}}_A$ but also to the difference between $\varepsilon_{m}$ and $\varepsilon_{s}$. 
	
	Combining Eqs.(\ref{eq:p4a}), (\ref{eq:E4b}) and (\ref{eq:p4c}), we can obtain
	\iffalse 	 
	\begin{equation}
		\begin{split}
			{\mathop{p}\limits^{\rightharpoonup}}_A=\frac{\varepsilon_{ p}-\varepsilon_{ m}}{\rvert\varepsilon_{ p}-\varepsilon_{ m}\rvert}4\pi\varepsilon_{m}\varepsilon_{0}{R_A^{*}}^{3}({\mathop{E}\limits^{\rightharpoonup}}_0+\frac{2{\mathop{p}\limits^{\rightharpoonup}}_B}{4\pi\varepsilon_{m}\varepsilon_{0}(2L)^{3}})
		\end{split},
		\label{eq:p31}
	\end{equation}	
	\begin{equation}
		\begin{split}
			{\mathop{p}\limits^{\rightharpoonup}}_A=\frac{\varepsilon_{ p}-\varepsilon_{ m}}{\rvert\varepsilon_{ p}-\varepsilon_{ m}\rvert}4\pi\varepsilon_{m}\varepsilon_{0}{R_A^{*}}^{3}({\mathop{E}\limits^{\rightharpoonup}}_0+\frac{-2\frac{\varepsilon _m-\varepsilon _s}{\varepsilon _m+\varepsilon _s} {\mathop{p}\limits^{\rightharpoonup}}_A}{4\pi\varepsilon_{m}\varepsilon_{0}(2L)^{3}})
		\end{split}.
		\label{eq:p32}
	\end{equation}
	With further derivation and simplification of Eq.(\ref{eq:p32}), we can deduce the expression for Eq.(\ref{eq:p33}),
	\fi
	\begin{equation}
		\begin{split}
			{\mathop{p}\limits^{\rightharpoonup}}_A = \frac{\frac{\varepsilon_p - \varepsilon_m}{|\varepsilon_p - \varepsilon_m|}  \left(4\pi\varepsilon_m\varepsilon_0{R_A^{*}}^{3}{\mathop{E}\limits^{\rightharpoonup}}_0\right)}{1 + \frac{\varepsilon_p - \varepsilon_m}{|\varepsilon_p - \varepsilon_m|}\left(2\frac{\varepsilon_m - \varepsilon_s}{\varepsilon_m + \varepsilon_s}  \frac{{R_A^{*}}^{3}}{(2L)^{3}}\right)} 
		\end{split}.
		\label{eq:p33}
	\end{equation}
	
	Eq.(\ref{eq:p33}) indicates that the direction of ${\mathop{p}\limits^{\rightharpoonup}}_A$ is still determined by the difference in dielectric constants of the particle and the solvent, although the presence of the substrate does influence its value. 
	
	The electrostatic interaction force $F$ between the dielectric particle and the planar substrate can be calculated using the obtained dipole moments. When considering the colinearity of the dipole moments ${\mathop{p}\limits^{\rightharpoonup}}_A$ and
	${\mathop{p}\limits^{\rightharpoonup}}_{A'}$, the electrostatic interaction force can be written as follows:
	\begin{equation}
		\begin{split}
			F& = -\frac{3p_A p_{A'} }{2\pi\varepsilon_m \varepsilon_0(2L)^{4}} \\
			&=\frac{\varepsilon _m-\varepsilon _s}{\varepsilon _m+\varepsilon _s}\frac{3{p_A}^2 }{2\pi\varepsilon_m \varepsilon_0(2L)^{4}} 
		\end{split}.
		\label{eq:FEDA}
	\end{equation}
	
	Eq.(\ref{eq:FEDA}) indicates that the electrostatic interaction force between the particle and the substrate is independent of the particle\('\)s permittivity. Instead, it is modulated by the difference in the dielectric constants of the substrate and the solvent. When ${\varepsilon_m > \varepsilon_s}$, a repulsive interaction is obtained, whereas an attractive interaction is produced when ${\varepsilon_m} < {\varepsilon_s}$.  
	
	\subsection*{MULTIPLE$-$IMAGE METHOD}  %{Multiple$-$Image Method}
	\label{MIM}
	
	In the MIM framework, the initial dipole moment induced by the uniform electric field ${\mathop{E}\limits^{\rightharpoonup}}_0$ in particle $A$ can be represented as a pair of effective point charges. These theoretical point charges are considered to be positioned at the  center of the particle, separated by a distance denoted as $l_0=10^{-4}{~\rm{nm}}$. In Fig.\ref{fig:Fig1Model}, the pair of point charges $q_{A,0}^{p}$ and $q_{A,0}^{m}$ in particle $A$ are illustrated as:
	
	\begin{subequations}
		\begin{align}
			&q_{A,0}^p=-q_{A,0}^{m}=\rvert{\mathop{p}\limits^{\rightharpoonup}}_{A,0}/{\mathop{l}\limits^{\rightharpoonup}}_0\rvert,\label{eq:mimq}\\
			&z_{A,0}^p=z_A+{l_0}/2, ~z_{A,0}^{m}=z_A-{l_0}/2,
			\label{eq:mimz}
		\end{align}
	\end{subequations}
	
	\noindent where the superscripts of $p$ and $m$ represent the positive and negative point charges, respectively.
	
	Based on the treatment for the inductance between a particle and a planar substrate, an image particle ${A'}$ is imaged at the position of $z_{A'}=-z_A$ with same radius $R_{A'}^{*}=R_A^{*}$. Within particle ${A'}$, two image charges $q_{{A'},0}^{m}$ and $q_{{A'},0}^{p}$ are symmetrically created at a distance $l_0$ from each other. For instance, considering $q_{A',0}^{m}$, it is produced at the position $z_{{A'},0}^{m}=-z_{A,0}^m$ with a charge of $q_{{A'},0}^{m}= \frac{\varepsilon _m-\varepsilon _s}{\varepsilon _m+\varepsilon _s} q_{A,0}^{m}$. 
	This results in an electrostatic interaction force between the dielectric particle $A$ and the planar substrate that is analogous to that between the reduced particle and the image particle ${A'}$. 
	The MIM is then employed to calculate the polarization of particle $A$ due to the image charges in particle ${A'}$.
	Taking the image charge $q_{A',0}^{m}$ as an example, its image charge $q_{A,1}^{m}$ in particle $A$ is at a distance of $r_{A,1}^{m}={R_{A}^*}^{2}/(z_A-z_{{A'},0}^{m})$ from the particle center, 
	with a charge of $q_{{A},1}^{m}=-\frac{\varepsilon_{ p}-\varepsilon_{ m}}{\rvert\varepsilon_{ p}-\varepsilon_{ m}\rvert}R_{A}^{*}q_{{A'},0}^{m}/(z_A-z_{{A'},0}^{m})$. 
	Similar procedures are repeated for $q_{A',0}^{p}$ and $q_{A,1}^{p}$ is generated in particle $A$. 
    Then the first group of image charges in particle $A$ is established after the first iteration of MIM. 
	Next, the MIM iteration progresses, generating successive groups of image charges in sequence. The magnitudes and positions of these image charges are determined based on the iteration step, 
	
	\begin{subequations}
		\begin{align}
			&q^{\lambda}_{{A'},n}=\frac{\varepsilon _m-\varepsilon _s}{\varepsilon _m+\varepsilon _s} q_{A,n}^{\lambda},
			\label{eq:mim1}\\
			&z_{{A'},n}^{\lambda}=-z_{A,n}^{\lambda},
			\label{eq:mim2}\\
			&q^{\lambda}_{A,n+1}=-\frac{\varepsilon_{ p}-\varepsilon_{ m}}{\rvert\varepsilon_{ p}-\varepsilon_{ m}\rvert}R_{A}^{*}q_{{A'},n}^{\lambda}/(z_A-z_{{A'},n}^{\lambda}),
			\label{eq:mim3}\\
			&r_{A,n+1}^{\lambda}={R_{A}^*}^{2}/(z_A-z_{{A'},n}^{\lambda}),
			\label{eq:mim4}\\
			&z^{\lambda}_{A,n+1}=z_A -r_{A,n+1}^{\lambda},
			\label{eq:mim5}
		\end{align}
	\end{subequations}
	where $\lambda=m,p$ and $n=0,1,\cdots$. %The result of Eq.\ref{eq:mim1} is consistent with that of Eq.\ref{eq:FEDA}. %When $\varepsilon_m > \varepsilon_s$, 
	
	When a sufficient iteration number of the MIM is reached, it becomes necessary to introduce the first compensation point charge, denoted as $q^{c_0}_{A,0}$, located at the center of particle $A$. The specific compensation charge is essential for preserving the conservation of charge within the particle,
	\begin{equation}
		q^{c_0}_{A,0}=-\sum_{n=0}^{N}q^{m}_{A,n}-\sum_{n=0}^{N}q^{p}_{A,n}.
		\label{eq:mim6}
	\end{equation}
	
	A similar process of image iteration is also required for the compensation charge $q^{c_0}_{A,0}$. This results in the generation of a series of corresponding image charges $q^{c_0}_{A,n'}$ and $q^{c_0}_{{A'},n'}$. Subsequently, the compensation point charges are once again introduced at the center of the particle. These compensation charges, along with their corresponding series of image charges, satisfy the following relationships:
	
	\begin{equation}
		q^{c_1}_{A,0}=-\sum_{n'=1}^{N'}{q^{c_0}_{A,n'}},
		\label{eq:mim7}
	\end{equation}
	\noindent where $n'$ and $N'$ are the index and the total number of iteration for the compensation charges. As the iteration process continues, the compensation charge sequences, denoted as $q^{c_2}_{A,0}, q^{c_3}_{A,0}...$, are generated. It has been observed that these sequences tend to approach zero, indicating that the compensation charges become increasingly insignificant as the iterations advance.
	
	Once the polarization, image charges, and compensation charges have been determined, the spatial potential $U^{q}$ outside particle $A$ and above the planar substrate can be precisely calculated. 
	Additionally, the electrostatic interaction force $F_A$ acting on particle $A$ from the substrate can be calculated using the acquired charges.
	
	\begin{equation}
		\begin{array}{ll}
			U^{q} = \frac{1}{4{\pi}\varepsilon_{m}{\varepsilon_{0}}}\sum\limits_{i=A,{A'}}
			\sum\limits_{\lambda=p,m,\atop c_{\tiny 0},c_{\tiny 1},c_{\tiny 2}...}
			\sum\limits_{n=0}^{N}
			\frac{q_{i,n}^{\lambda}}{[(x_{i,n}^{\lambda}-x)^2
				+(y_{i,n}^{\lambda}-y)^2+(z_{i,n}^{\lambda}-z)^2]^{1/2}}
		\end{array},
		\label{eq:19}
	\end{equation}
	\begin{equation}
		\begin{array}{ll}
			F_A = \frac{1}{4{\pi}{\varepsilon_{m}}{\varepsilon_{0}}}
			\sum\limits_{\lambda_{_A},\lambda_{_{A'}}=p,m,\atop c_0,c_1,c_2...}
			\sum\limits_{n_{_A},n_{_{A'}}=0}^{N}
			\frac{q_{A,n_A}^{\lambda_A}   q_{{A'},n_{A'}}^{\lambda_{A'}}}
			{(x_{A,n_A}^{\lambda_A}-x_{{A'},n_{A'}}^{\lambda_{A'}})^2
				+(y_{A,n_A}^{\lambda_A}-y_{{A'},n_{A'}}^{\lambda_{A'}})^2+(z_{A,n_A}^{\lambda_A}-z_{{A'},n_{A'}}^{\lambda_{A'}})^2}
		\end{array},
		\label{eq:20}
	\end{equation}
	
	\noindent where $x_{i,n}^{\lambda}=y_{i,n}^{\lambda}=0$. In all calculations, a total of $N= 400$ iterations are employed to ensure numerical convergence. %It is worth emphasizing that does not consider the impact of the surface charge density $\sigma$ on the substrate.
	Additionally, it is crucial to acknowledge that the particles are  electrically neutral, and therefore, it does not undergo any net force in an external uniform field ${\mathop{E}\limits^{\rightharpoonup}}_0$.
	
	\section*{RESULTS AND DISCUSSIONS}  %{Results and Discussions}
	\label{Results}
	
\begin{figure*}[htbp]
		\centering
		\includegraphics[width=0.99\textwidth,trim=0 0 0 0,clip]{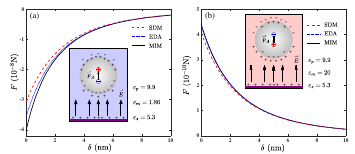}
		\caption{(color online) Dependence of the electrostatic interaction force $F$ between a neutral dielectric particle and a planar substrate.
			%The separation gap is set to $\delta=5~{\rm nm}$. 
			The radius of the particle $A$ is $R_{A}={10~\rm nm}$ and its relative permittivity is set to $\varepsilon_p=9.9$. The charge density of the planar substrate is $\sigma=1\rm~e\cdot nm^{-2}$ and its relative permittivity is $\varepsilon_s=5.3$. The relative permittivities of the solvent are (a) $\varepsilon_m=1.86$ and (b) $\varepsilon_m=20.0$. Solid, dotted and dashed lines represent the results using MIM, EDA, and SDM, respectively. The insets provide a  schematic illustration of the multiple dipole moment.}%the spatial arrangement of bound charges around particle surface.}
	\label{Fig2FDelta}
\end{figure*}

	Fig.\ref{Fig2FDelta} shows the electrostatic interaction force between a neutral dielectric particle and a dielectric planar substrate using the MIM and the EDA. In accordance with parameters from previous experiments\cite{Verveniotis2011}, the particle is made of alumina with a relative permittivity $\varepsilon_p=9.9$. The planar substrate, which features a charged nanodiamond surface with a net charge density of $\sigma=1\rm~e\cdot nm^{-2}$, also has a relative permittivity of $\varepsilon_s=5.3$. Additionally, in Fig.\ref{Fig2FDelta}(a), 
	the medium is an insulating fluorocarbon solution with a relative permittivity of $\varepsilon_m=1.86$. Notably, in Fig.\ref{Fig2FDelta}(b), an alternative configuration is presented, utilizing acetone as the medium with a higher dielectric constant of $\varepsilon_m=20.0$\cite{Lindgren2018a}. In this study, the selected particle radius is set at $10~{\rm nm}$, a common size for nanoscale particles.
	
	The calculations also involve the use of the simple$-$dipole method (SDM), which considers only the charges $q_{A,0}^{\lambda}$ of the particle and $q_{{A'},0}^{\lambda}$ of the substrate. 
	The results obtained from all three methods (MIM, EDA, and SDM) exhibit agreement for large separations $\delta$. However, for small separations, the absolute values of the calculated forces obtained from the MIM are greater than those from the EDA and SDM. This discrepancy becomes more pronounced as the separation decreases. The underlying reason for this difference lies in the distinct approaches employed by these methods. In the SDM, the multiple mutual polarizations between the particle and the substrate are not considered. In the EDA, the polarized dipole moment is positioned at the center of the particle $A$, whereas in the MIM, the image charges of the dipole moment are dispersed away from the center of the particle. This implies that the MIM approach captures the intricate interactions more accurately, thus providing a more comprehensive understanding of the system.
	
In Fig.\ref{Fig2FDelta}, negative and positive values of $F$ represent attractive and repulsive interaction forces, respectively. The relative permittivities of the particle and the substrate are fixed at ${\varepsilon_p}=9.9$ and $5.3$, respectively. %The separation gap is set to $\delta=5~{\rm nm}$.
The results obtained from all three methods indicate an attractive interaction in a weakly polarizable solvent, $\varepsilon_m=1.86$, shown in Fig.\ref{Fig2FDelta}(a) and a repulsive force in a strongly polarizable solvent, $\varepsilon_m=20$ shown in Fig.\ref{Fig2FDelta}(b). This reversed polarization is in agreement with the predictions of EDA and the previous work of dielectrophoresis (DEP)\cite{Lindgren2018a, Cetin2011}.
When the relative permittivity of the particle is higher than that of the solvent, $\varepsilon_p > \varepsilon_m$, more polarization charges accumulate at the particle side, schematically indicated by ${\mathop{p}\limits^{\rightharpoonup}}_A$ in Fig.\ref{Fig2FDelta}(a).
The direction of ${\mathop{p}\limits^{\rightharpoonup}}_A$ is aligned the direction of the external electric field ${\mathop{E}\limits^{\rightharpoonup}}_0$. This polarization direction of the particles is consistent with the result shown in Eq.(\ref{eq:p2}). On the contrary, when the solvent has a higher relative permittivity $\varepsilon_m>\varepsilon_p$, it becomes more polarizable than the particle. The reversed polarization charges appear at the solvent interface. In this case, the resultant dipole moment is depicted in the inset of Fig.\ref{Fig2FDelta}(b), which is in the oppositve polarization direction to the external electric field ${\mathop{E}\limits^{\rightharpoonup}}_0$.

\begin{figure*}[htbp]
	\centering
	\includegraphics[width=0.99\textwidth,trim=5 0 0 0,clip]{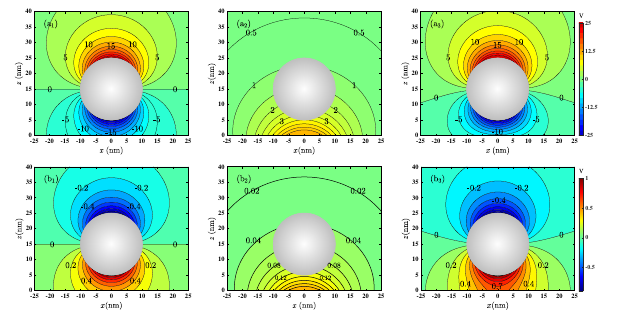}
	\caption{(color online) Spatial distribution of electric isopotential $U^{q}$ of the polarized charges. The same parameters are used as in Fig.\ref{Fig2FDelta}. The separation gap is $\delta=5\rm nm$. The relative permittivities of the solvent are (a) $\varepsilon_m=1.86$ and (b) $\varepsilon_m=20.0$, respectively.}
	\label{FigUDist}	
\end{figure*}

The above results have shown that the attractive and repulsive interaction between a neutral dielectric particle and a planar dielectric substrate can be entirely reversed by adjusting the dielectric constant of the solvent. It is of significant importance to discover the influence of polarized charges on spatial distribution of electric isopotential $U^q$. Fig.\ref{FigUDist} plots the $U^q$ outside of the particle calculated by the MIM with zero electric potential at infinity. % under the same conditions as in Fig.\ref{Fig2FDelta}. 
Fig.\ref{FigUDist}$(\rm a_1)$ plots the $U^q$ generated solely by the polarized charges on particle $A$, with same parameters as shown in Fig.\ref{Fig2FDelta}(a).
The initial uniform electric field is expected to be disrupted around particle $A$. 
An axial symmetry with the same $U^q$ in the $x$ direction is observed due to the identical nature of the particle. In the $z$ direction, an almost axial symmetry with reversed $U^q$ is also observed though a tiny difference exists. Positive and negative $U^q$ appear around the upper and lower regions of particle $A$, respectively. This is equivalent to a dipole moment aligned with the same direction of ${\mathop{E}\limits^{\rightharpoonup}}_0$, consistent with the result of Eq.(\ref{eq:p33}). 
In Fig.\ref{FigUDist}$(\rm a_2)$, the positive $U^q$ generated by the polarized charges on the substrate indicates that the polarized charges on the substrate are positive. In the other words, it is actually calculated by the dipole moment at  particle ${A'}$. This positive $U^q$ above the substrate implies that the direction of ${\mathop{P}\limits^{\rightharpoonup}}_{A'}$ aligns with the direction of ${\mathop{E}\limits^{\rightharpoonup}}_0$. 
Furthermore, the nonuniform distribution of $U^q$
indicates the generation of a nonuniform electric field, with stronger and weaker electric fields occurring at the lower and upper regions of particle $A$. This gradient electric field directly results in the occurrence of positive dielectrophoresis. Additionally, it leads to an attractive electrostatic interaction between particle $A$ and the substrate. 
Fig.\ref{FigUDist}$(\rm a_3)$ plots the combined $U^q$ of the polarized charges on the particle $A$ and the substrate, as shown in Fig.\ref{FigUDist}$(\rm a_1)$ and Fig.\ref{FigUDist}$(\rm a_2)$. 
It is evident that the former significantly dominates the distribution of $U^q$, with positive and negative $U^q$ still appearing around the upper and lower parts regions of particle $A$, respectively. 

We also plot the distribution of $U^q$ for the case of Fig.\ref{Fig2FDelta}(b) as shown in Fig.\ref{FigUDist}. The solvent has a large relative permittivity, $\varepsilon_m=20$. All similar panels in Fig.\ref{FigUDist} exhibit analogous  properties of $U^q$.
In the $x$ direction, the identical nature of particle results in complete symmetry, as shown in Fig.\ref{FigUDist}($\rm b_1$). Conversely, in the $z$ direction, the dipole moment at particle $A$ is reversed as expected from Eq.(\ref{eq:p33}), producing positive and negative $U^q$ at the lower and upper regions of particle $A$, respectively. 
As shown in Fig.\ref{FigUDist}($\rm b_2$), positive $U^q$ is expected to appear at the region above the substrate, indicating a repulsive electrostatic interaction between particle $A$ and the substrate. Likewise, $U^q$ in Fig.\ref{FigUDist}($\rm b_3$) is predominantly influenced by the polarization charges at the particle $A$.

\begin{figure*}[htbp]
	\centering
	\includegraphics[width=0.95\textwidth,trim=130 410 100 130,clip]{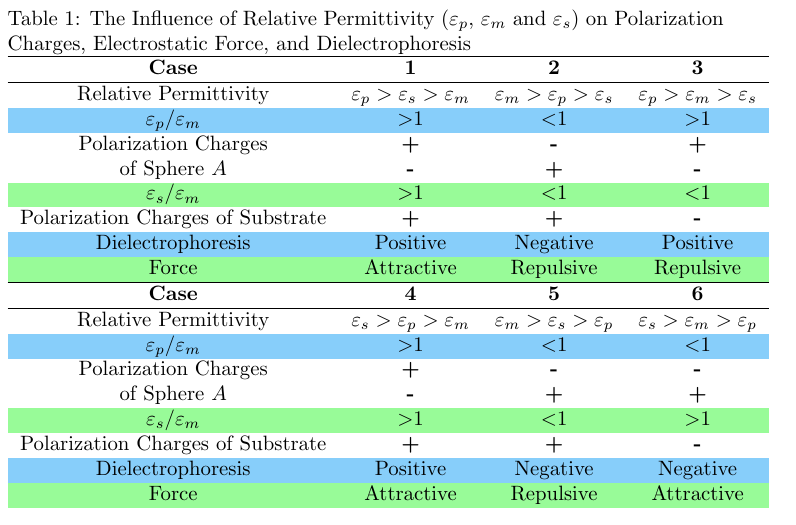}
	%\caption{(color online) Spatial distribution of electric isopotential $U^{q}$ of the polarized charges. The same parameters are used as in Fig.\ref{Fig2FDelta}. The separation gap is $\delta=5\rm nm$. The relative permittivities of the solvent are (a) $\varepsilon_m=1.86$ and (b) $\varepsilon_m=20.0$m, respectively.}
	\label{Table1}	
\end{figure*}

In order to systematically investigate the influence of relative permittivity ($\varepsilon_{p}$,  $\varepsilon_{m}$ and $\varepsilon_{s}$) on the interaction force, the numerical results are summarized in Table 1, as predicted by Eq.(\ref{eq:p2}), Eq.(\ref{eq:p4c}) and Eq.(\ref{eq:p33}). 
As indicated by Eq.(\ref{eq:p2}), for ${\varepsilon_{p}} > {\varepsilon_{m}}$, the direction of ${\mathop{p}\limits^{\rightharpoonup}}_A$ is same to that of ${\mathop{E}\limits^{\rightharpoonup}}_0$. Conversely, the direction of ${\mathop{p}\limits^{\rightharpoonup}}_A$ is reversed arises when ${\varepsilon_{p}} < {\varepsilon_{m}}$. This leads to the generation of positive and negative dielectrophoresis depending on the polarization direction of particle $A$. In Eq.(\ref{eq:p4c}), the polarization of the image particle ${A'}$ is determined by the difference between ${\varepsilon_{s}}$ and ${\varepsilon_{m}}$ along with the polarization state of the particle $A$. However, according to Eq.(\ref{eq:p33}), the electrostatic interaction force between particle $A$ and the substrate solely depends on the difference between ${\varepsilon_{s}}$ and ${\varepsilon_{m}}$. The simulation results obtained using the MIM validate the theoretical predictions of the EDA. 
Taking case $1$ for example, when ${\varepsilon_{p}} > {\varepsilon_{s}} > {\varepsilon_{m}}$, it corresponds to the results shown in Fig.\ref{Fig2FDelta}(a) and Fig.\ref{FigUDist}($\rm a$). 
In this scenario, particle $A$ exhibits a polarization direction consistent with the direction of  ${\mathop{E}\limits^{\rightharpoonup}}_0$, resulting in positive dielectrophoresis. As a consequence, positive polarized charges accumulate on the surface of the substrate, leading to an attractive electrostatic interaction force between particle $A$ and the substrate.

\begin{figure*}[htbp]
	\centering
	\includegraphics[width=0.99\textwidth,trim=0 0 0 0,clip]{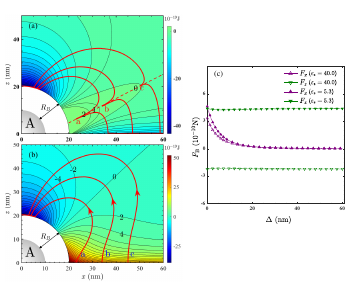}
	\caption{(color online) Spatial distribution of external interaction energy on particle $B$ with particle $A$ and the substrate in the medium with $\varepsilon_m=20.0$. particle $A$ is fixed on the substrate with a separation distance of $\delta=0~\rm nm$, and the substrate is located at $z=-10~\rm nm$. The relative permittivities of the substrate are (a) $\varepsilon_s=40.0$ and (b) $\varepsilon_s=5.3$, respectively. The solid red lines represent gradient lines, and the width of the white sector corresponds to the radius of particle $B$. 
	(c) The electrostatic force $F_B$ as a function of the gap $\Delta$ between particles $A$ and $B$.
	Upward triangles and downward triangles denote $F_x$ and $F_z$, respectively. Open and solid symbols represent the results of (a) $\varepsilon_s=40.0$ and (b) $\varepsilon_s=5.3$, respectively.
	}
	\label{Fig:Fig4Fdeltat}
\end{figure*}

Since the electrostatic interaction force between particle $A$ and the substrate is directly modulated by the difference of the relative permittivity between the solvent and the substrate, it is interesting to explore how the relative permittivity of the substrate influences the self$-$assembly of the particles. Let us consider the case of two identical particles $A$ and $B$. particle $A$ is fixed on the substrate, and particle $B$ is movable. The electrostatic interaction force between the particles $A$ and $B$ is calculated using the MIM\cite{Li2022,Li2024}, as well as those between the particles and the substrate.
For particle $B$, the electrostatic interaction force is composed of two parts. The first part is the external interaction forces between the charges in particle $B$ and the charges in particle $A$ and the substrate. The second part involves the internal interaction forces among the internal charges within particle $B$. Consquently, the former dominates the motion of particle $B$. In Fig.\ref{Fig:Fig4Fdeltat}(a), the distribution of the external interaction energy on particle $B$ from particle $A$ and the substrate are plotted for the relative permittivities of the substrate, the solvent and the particle are set to $\varepsilon_s=40.0$, $\varepsilon_m=20.0$ and $\varepsilon_p=9.9$, corresponding to the case $6$ in Table 1.
Under these conditions, the electrostatic interaction force between the substrate and each particle is attractive.
The three red lines represent typical gradient lines.
The direction of the tangent at any point on a gradient line indicates the electrostatic interaction force of particle $B$ from particle $A$ and the substrate.
It is evident that lower energy levels occur on the upper side of particle $A$ and at the surface of the substrate.
A ridge line for the electrostatic energy, denoted by the dashed line, is clearly observed, dividing the space into two regions. This ridge line does not intersect with the gradient lines. 
If particle $B$ falls exactly on the ridge line, such as the points ${\rm a}$, $b$ or $\rm c$, the resultant force on particle $B$ will push it away along the dashed line. 
If particle $B$ deviates slightly from the ridge line, 
it will move toward the upper side of particle $A$ or the substrate, respectively.    
When the relative permittivity of the substrate is $\varepsilon_s=5.3$, the 
external interaction energy on particle $B$ and the corresponding gradient lines are also plotted in Fig.\ref{Fig:Fig4Fdeltat}(b). Lower energy levels occur at the upper side of particle $A$, while the higher energy levels appear at the surface of the substrate, respectively. Notably, no ridge line appears. These characteristics indicate that particles $A$ and $B$ will be aligned along a line in the direction of ${\mathop{E}\limits^{\rightharpoonup}}_0$. 

In Fig.\ref{Fig:Fig4Fdeltat}(c), the electrostatic force of particle $B$ received from particle $A$ and the substrate is plotted when particle $B$ is located at the surface of the substrate. It is noticeable  that $F_x$ keeps positive, indicating a repulsive force, regardless of the relative permittivity of the substrate. This consistent behavior can be attributed to the similar nature of particles of $A$ and $B$, which result in parallel polarizations. However, for large $\varepsilon_s=40$ and small $\varepsilon_s=5.3$, we observe negative and positive values for $F_z$, indicating attractive and repulsive forces, respectively. These observations align with the results illustrated in Figs.\ref{Fig:Fig4Fdeltat}(a) and \ref{Fig:Fig4Fdeltat}(b). 

\begin{figure*}[htbp]
	\centering
	\includegraphics[width=0.75\textwidth,trim=160 160 160 170,clip]{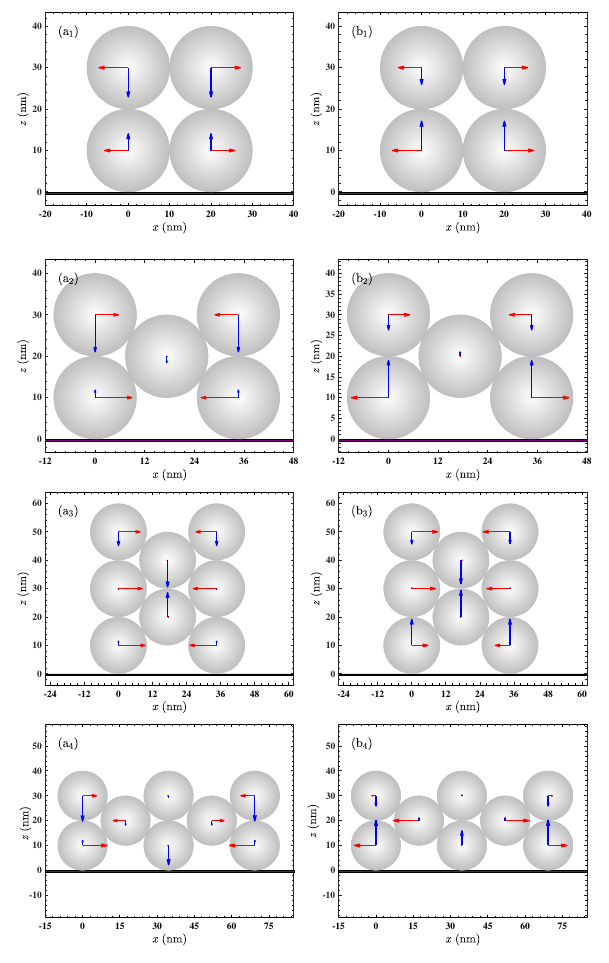}
	\caption{(color online) The electrostatic forces experienced by particles with different numbers and structures with (a) $\varepsilon_s=40.0$ and (b) $\varepsilon_s=5.3$. The vertical axis (denoted as z) is indicated by blue vectors, and the horizontal axis (denoted as x) is indicated by red vectors. Within different subfigures, the lengths of vectors in different colors are adaptive in order to more effectively demonstrate and comprehend electrostatic forces.}
	\label{Fig:GeoStruct}
\end{figure*}

When more particles are introduced, it becomes essential to observe the stable geometrical structure that the particles will aggregate under the influence of an external electric field. Fig.\ref{Fig:GeoStruct} employs the same parameters as Fig.\ref{Fig:Fig4Fdeltat} with the dielectric constants set at $\varepsilon_s=40.0$ and $\varepsilon_s=5.3$ for the left and right panels of Fig.\ref{Fig:GeoStruct}, respectively. Two distinct geometrical structures are examined: the square structure and the hexagonal structure. 
In Fig.\ref{Fig:GeoStruct}, the length of the arrows represents the magnitude of the electric force in the $x$ and $z$ directions.
Generally, the characteristics of the electrostatic force between the substrate and the cluster exhibit similarities to those between a single particle and the substrate in the $z$ direction. When $\varepsilon_s=40.0$, the resultant electrostatic force between the cluster and the substrate is attractive, while it becomes repulsive when $\varepsilon_s=5.3$. Additionally,  the electrostatic force remains consistently attractive among the particles, suggesting that the chain-like structure is stable for the particles in the $z$ direction.
However, the stability of particle aggregation in the $x$ direction is closely associated with the geometrical structure. 
For instance, in Fig.\ref{Fig:GeoStruct}$\rm (a_1)$ and Fig.\ref{Fig:GeoStruct}$\rm(b_1)$, the square arrangement of particles is unstable because the electrostatic force always points outward from the interior of the cluster in the $x$ direction. Conversely, the hexagonal structure exhibits stability in the $x$ direction since the peripheral particles experience inward electrostatic forces for $\varepsilon_s=40$, as depicted in Fig.\ref{Fig:GeoStruct}$\rm (a_2)$. 
In contrast, when $\varepsilon_s=5.3$, the particles tend to separate in the $x$ direction due to outward electrostatic forces. 
Upon introducing an additional layer of particles in the $z$ direction, the hexagonal structure proves to be stable for both both $\varepsilon_s=40$ and $\varepsilon_s=5.3$, as shown in Fig.\ref{Fig:GeoStruct}$\rm (a_3)$ and Fig.\ref{Fig:GeoStruct}$\rm (b_3)$. However, adding an extra layer of particles in the $x$ direction results in stability within the hexagonal structure for $\varepsilon_s=40$, as shown in Fig.\ref{Fig:GeoStruct}$\rm (a_4)$. On the other hand, for $\varepsilon_s=5.3$, the hexagonal structure becomes unstable as the electrostatic force on the particles points outward, as depicted Fig.\ref{Fig:GeoStruct}$\rm (b_4)$.
%\newpage

\section*{CONCLUSIONS}  %{Conclusions}
\label{Conclusions}

In conclusion, we conducted numerical investigations of the electrostatic interaction forces between dielectric spherical particles and a uniformly charged substrate using the EDA and the MIM. By applying the CM relation, an equivalent conducting spherical particle with a smaller radius and a polarized dipole moment was introduced in a uniform electrostatic field. 
Image charges were symmetrically placed in the substrate depending on the dielectric constants of the solvent medium and the substrate. 
In the EDA, an analytical formula was derived, while the computation iterations were established in the MIM for analyzing the electrostatic interactions thoroughly.

Both the EDA and the MIM results initially revealed the attractive and repulsive electrostatic interactions between the neutral particle and the substrate for $\varepsilon_s>\varepsilon_m$ and $\varepsilon_s<\varepsilon_m$, respectively. The analysis of the distribution of electrostatic potential indicated that the variations of the dielectric constants of the particle and the solvent medium can lead to a reversal of polarization direction of the particle. The polarized charges of the substrate consistently created a nonuniform local electrostatic potential around the particle, resulting in both attractive and repulsive electrostatic interactions. Generally, a polarization direction that is the same as the external electric field occurs when $\varepsilon_p/\varepsilon_m>1$, whereas an opposite polarization arises for $\varepsilon_p/\varepsilon_m<1$.
Furthermore, attractive and repulsive electrostatic interactions are observed for $\varepsilon_s/\varepsilon_m>1$ and $\varepsilon_s/\varepsilon_m<1$, respectively.
Subsequent analysis of the electrostatic energy and its gradient for a system with two particles indicates that the electrostatic field of the substrate arranges particles in a line.
Notably, higher dielectric constants of the substrate tend to cause neutral particles to absorb to the substrate, whereas lower dielectric constants push neutral particles away from the substrate. When multiple particles are considered, they exhibit a preference for forming a hexagonal structure rather than a square one. These findings offer valuable insights into the electrostatic manipulation of the design and fabrication of advanced materials.

\appendix
\section{THE IMAGE OF A POINT CHARGE}  %{The Image of a Point Charge }
\label{Appendix A}
\begin{figure}[htbp]
	\centering
	\includegraphics[width=0.45\textwidth,trim=185 345 130 290,clip]{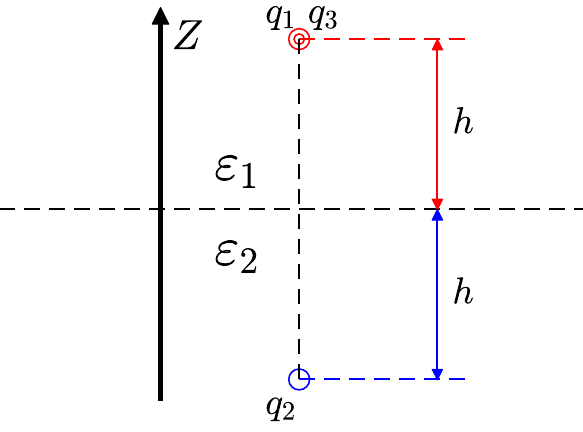}
	\setcounter{figure}{0}
	\caption{(color online) Schematic of the image method for a point charge in proximity to  the interface of two dielectrics, where the dielectric constants are $\varepsilon_1$ and $\varepsilon_2$, respectively. $h$ is the distance from the point charge to the interface.}
	\label{fig:FigAppModel}
\end{figure}
In the computation of electric potential within either dielectric $1$ or $2$, we employ image charges situated in the alternate dielectric medium as a substitute for polarization charges at the interface. Correspondingly, we consider the entire space as a homogeneous medium filled with a dielectric constant of $\varepsilon_1$ for dielectric $1$, and $\varepsilon_2$ for dielectric $2$, as shown in Fig. \ref{fig:FigAppModel}. 

The potential at any point in the space can be expressed as
\begin{equation}
	\varphi \left ( x,y,z \right ) = 
	\begin{cases} 
		\frac{1}{4\pi \varepsilon _1} \left [ \frac{q_1}{\sqrt{x^2+y^2+\left ( z-h \right )^2 } } +\frac{q_2}{\sqrt{x^2+y^2+\left ( z+h \right )^2 } } \right ], & \text{if } z\ge 0 \\
		\frac{1}{4\pi \varepsilon _2}\frac{q_1+q_3}{\sqrt{x^2+y^2+\left ( z-h \right )^2 } }. & \text{if } z\le  0 \label{eq: electric potential}  
	\end{cases}
\end{equation}
According to the boundary conditions:
\begin{subequations}
	\begin{align}
		&\varphi _{1}|_{z=0}=\varphi _{2}|_{z=0},\label{eq:electric potential1}\\
		&\varepsilon_{1}\frac{\partial\varphi _1}{\partial z} \Bigr| _{z=0}=\varepsilon_{2}\frac{\partial\varphi _2}{\partial z} \Bigr| _{z=0},\label{eq:electric potential2}
	\end{align}
\end{subequations}
we can obtain
\begin{subequations} 
	\begin{align}     
		&\frac{1}{\varepsilon _1}\left ( q_1+q_2 \right )=\frac{1}{\varepsilon _1}\left ( q_1+q_3 \right ),  \label{eq:electric potentialq1}  \\    
		&      q_1-q_2=q_1+q_3. \label{eq:electric potentialq2}
	\end{align}
\end{subequations}
Upon simplification of Eqs.(\ref{eq:electric potentialq1}) and (\ref{eq:electric potentialq2}), Eqs.(\ref{eq:electric potentialq3}) and (\ref{eq:electric potentialq4}) are derived,
\begin{subequations} 
	\begin{align}     
		& q_2= \frac{\varepsilon _1-\varepsilon _2}{\varepsilon _1+\varepsilon _2} q_1,\label{eq:electric potentialq3}  \\    
		& q_3=-\frac{\varepsilon _1-\varepsilon _2}{\varepsilon _1+\varepsilon _2} q_1. \label{eq:electric potentialq4}
	\end{align}
\end{subequations}

For the image of an electric dipole near a boundary interface, both magnitude and direction need to be considered. Specifically, when the direction of the original dipole is perpendicular to the interface, 
the dipole moment of the image dipole, denoted as ${\mathop{p}\limits^{\rightharpoonup}}_{A'}$, is proportional to the original dipole moment, denoted as ${\mathop{p}\limits^{\rightharpoonup}}_A$. The magnitude and direction can be determined by the following equation:
\begin{equation}
	\begin{split}
		{\mathop{p}\limits^{\rightharpoonup}}_{A'}=-\frac{\varepsilon _1-\varepsilon _2}{\varepsilon _1+\varepsilon _2} {\mathop{p}\limits^{\rightharpoonup}}_A
	\end{split}.
	\label{eq:electric potentialp}
\end{equation}

\section*{Acknowledgments}
This work is financially supported by the National Natural Science Foundation of China (Grant No. 11574153) and the foundation of the Ministry of Industry and Information Technology of China (Grant No. TSXK2022D007).

\end{document}